# Cerebral Aneurysm Flow Diverter Modeled as a Thin Inhomogeneous Porous Medium in Hemodynamic Simulations

Armin Abdehkakha MS[1,3], Adam L. Hammond MS[1,3], Tatsat R. Patel MS[1,3], Adnan H. Siddiqui MD PhD[3-5], Gary Dargush PhD[1], Hui Meng PhD[1-3]

[1]Department of Mechanical & Aerospace Engineering at the University at Buffalo, Buffalo, New York, USA

[2]Department of Biomedical Engineering at the University at Buffalo, Buffalo, New York USA

[3]Canon Stroke and Vascular Research Center at the University at Buffalo, Buffalo, New York USA

[4]Department of Neurosurgery, Jacobs School of Medicine at the University at Buffalo, Buffalo New York, USA

[5]Department of Radiology, Jacobs School of Medicine at the University at Buffalo, Buffalo New York, USA

## ABSTRACT

Rapid and accurate simulation of cerebral aneurysm flow modifications by flow diverters (FDs) can help improving patient-specific intervention and predicting treatment outcome. However, with explicit FD devices being placed in patient-specific aneurysm model, the computational domain must be resolved around the thin stent wires, leading to high computational cost in computational fluid dynamics (CFD). Classic homogeneous porous medium (PM) methods cannot accurately predict the post-stenting aneurysmal flow field due to the inhomogeneous FD wire distributions on anatomic arteries. We propose a novel approach that models the FD flow modification as a thin inhomogeneous porous medium (iPM). It improves over classic PM approaches in that, first, FD is treated as a screen, which is more accurate than the classic Darcy–Forchheimer relation based on 3D PM; second, the pressure drop is calculated using local FD geometric parameters across an *inhomogeneous* PM, which is more realistic. To test its accuracy and speed, we applied the iPM technique to simulate the post-stenting flow field in three patient-specific aneurysms and compared the results against CFD simulations with explicit FD devices. The iPM CFD ran 500% faster than the explicit CFD while achieving 94%-99% accuracy. Thus iPM is a promising clinical bedside modeling tool to assist endovascular interventions with FD and stents.

# INTRODUCTION

Intracranial aneurysms (IAs) result from pathological remodeling of cerebral blood vessel walls [1]. If left untreated, an IA can rupture and cause fatal subarachnoid hemorrhage. To prevent IA rupture, endovascular devices such as flow diverters (FDs) and embolic coils can be deployed to induce aneurysm thrombosis and thus occlusion [2]. The FD is an especially valuable option for treating large, wide-necked and complex aneurysms that are otherwise untreatable [3]. Being essentially a densely meshed stent, a flow diverter is a highly flexible, self-expanding porous tubular mesh of individual wires (30–50 $\mu m$ diameter) braided into a thin, single or multi-layered screens with small pores (~100 $\mu m$) [2]. When implanted across the aneurysm orifice, an FD functions to divert most of the blood flow away from the aneurysm so as to induce thrombotic occlusion and parent-vessel reconstruction. Treatment outcome is highly dependent on aneurysmal hemodynamic factors [4]. Since the post-FD hemodynamics can indicate the likelihood of aneurysm healing [4], it is highly desirable to model the FD intervention procedure *a priori* and predict the post-treatment aneurysmal flow, so as to improve patient-specific intervention outcome. This concept is known as *virtual intervention* [5].

Recently researchers have used computational fluid dynamics (CFD) simulations to numerically calculate flow field changes induced by FD implantation. They have correlated the computed hemodynamic factors with the aneurysm's healing outcome using machine-learning models, with encouraging results [4]. This shows the promising potential of using computation to assess treatment strategies and improve patient-specific treatment outcome. However, it is currently challenging to implement CFD simulations for FD treatment in the routine clinical workflow. While the virtual deployment of stents into computational models is now a mature technique and can be done within seconds [2], the post-stenting CFD comes with high computational cost. In the explicit presence of an FD device, modelling the fluid domain around the thin stent wires requires fine discretization [6], which increases the computational resource requirements for volumetric meshing and flow simulations [7]. The FD wires are typically extremely thin (~30 $\mu m$) compared to the base mesh size ($\approx 0.1\ mm$). To explicitly model the flow around these thin wires, the mesh size near the wires must be refined (typically to $\approx 10 \mu m$). This refinement accounts for the majority of mesh elements in the fluid domain, drastically increasing processing time compared to CFD on the untreated aneurysm. This, in turn, leads to high computational cost, with one CFD simulation typically taking ~24 hours on supercomputer clusters [8].

Researchers have addressed the computational cost issue by modeling the post-treatment flow in steady-state instead of pulsatile flow condition [9], or employing alternative meshing techniques such as body-fitted and immersed-body methods [10], in order to obtain the post-stenting hemodynamic results faster. These techniques have accelerated the computations to some extent. However, the flow discretization and CFD simulations are still prohibitively time-consuming, as long as the device is explicitly represented in the computational domain. This represents a barrier to clinical translation, since any bedside tool must be both accurate and fast enough to help the clinician plan the best treatment in actual endovascular interventions.

An efficient alternative to CFD with explicit FD is to represent the FD implicitly as a Porous Medium (PM), as demonstrated by the pioneering work by Augsburger et al. [8]. This can drastically reduce the computational costs of obtaining post-treatment hemodynamics, since CFD can be solved on coarse computational grids equivalent to the un-stented case. Augsburger et al. [8] implicitly represented an FD as a classic homogeneous PM based on Darcy–Forchheimer equation, $\frac{\Delta P}{\Delta L} = -(\frac{\mu}{K}u + C_2 \frac{1}{2}\rho u^2)$, where $\Delta P$ is pressure drop, $\Delta L$ is porous media thickness, $K$ is permeability, $C_2$ is inertial resistance factor, $u$ is flow velocity, $\rho$ is flow density, and $\mu$ is flow viscosity [11]. The coefficients $K$ and $C_2$ for specific FD meshes were measured empirically in a series of numerical experiments. Their results show that the accuracy of the PM-based CFD was less than 70% on average.

A possible cause of the low accuracy in the PM flow prediction in Augsburger et al. [8] is that the PM model based on the Darcy–Forchheimer equation was derived from the Hagen-Poiseuille equation for *fully-developed flow* in a pipe [12, 13]. A flow diverter consists of only a single thin layer of wires, with a thickness of $\sim 40\mu m$. Clearly, the fully-developed pipe flow assumption is not appropriate for a thin layer of pores (See Section 2.1). Another possible cause of the low accuracy is the homogeneous PM assumption.

Raschi et al. [14] recognized that a stent or FD is merely a thin layer, not a "3D" porous medium; hence instead of the Darcy-Forchheimer equation, they treated the FD as a flat screen with a pressure drop represented by a phenomenological equation in the handbook by Idelchik [15].

Being widely used in industry, this "Idelchik model" was first introduced to stent modeling and testing by Kim et al. [16]. Raschi et al. [14] further added correction factors to the viscous and inertial resistance coefficients of the Idelchik model. This modified model was later dubbed as the "Raschi model" [17]. In their numerical testing of the "Idelchik model" and the "Raschi model", Dazeo et al. showed the original "Idelchik model" still outperformed the "Raschi model" in accuracy [17].

It should be noted that all of the past approaches shown above assumed homogeneous PM. In reality, FDs implanted in real aneurysms have spatially varying pore sizes due to the complex anatomic geometry of a cerebral artery (see Section 2.1) and are thus inhomogeneous. Efforts have been made to improve the homogeneous PM models to account for ranges of braiding angles [18], stent orientation [19], and different types of FDs (i.e. PED, Silk+, and FRED) [20]; however, the recommended homogeneous parameters cannot extrapolate well to the general complex, inhomogeneous FD cases.

To account for the inhomogeneity effect of implanted FDs, Farsani et al. [6] modeled the FD as an inhomogeneous PM for the first time. They heterogenized at the scale of the untreated mesh cells by determining whether portions of the FD wires passed through individual tetrahedral mesh cells. For each mesh cell with wires passing through them, they applied a PM pressure drop based on the Darcy model ($\nabla P = -\frac{\mu u}{K}$) [21]. Their PM approach predicted hemodynamic parameters of 3 patient-specific IA cases, which agreed with explicit-device CFD simulation results. However, it is not clear if the results of Farsani et al. [6] are physical, for two reasons. First, the Darcy model used to acquire $\nabla P$ is derived by volume-averaging the collective effects of many pores on the fluid pressure drop over the porous medium [11]. Hence the $\nabla P$ prediction by the Darcy model is not valid as a means of deducing local pressure drop on the scale of the mesh element (much smaller than the scale of the pores) as was done in Farsani et al. [6]. Second, the Darcy model for PM pressure drop implicitly assumes that the thickness of the FD is so large that the flow through the pores is fully developed, which is not the case for a screen-like stent. Since their model of PM pressure drop across the stent layer is not appropriate, it is questionable if the accuracy results of Farsani et al. [6] would be repeatable in a larger cohort. Another concern is that the computational speed increase by the PM method of Farsani et al. [6] could be compromised by the higher computational cost incurred by the use of tetrahedral mesh cells that is required by their approach.

Compared to the more commonly used polyhedral mesh, simulations using a tetrahedral volume mesh typically required 10 to 15 times longer simulation runtime [22]. This limitation could entirely erode the benefits of using a PM approach in the first place.

We hypothesize that by adopting inhomogeneous PM properties to CFD of post-stenting hemodynamics, accuracy will be drastically increased. To facilitate *fast and accurate* CFD simulation of post-stenting aneurysmal hemodynamics, in this paper we present a novel *thin-walled, inhomogeneous PM* approach called iPM. We use the thin-screen model by Idelchik [15] to determine inhomogeneous PM properties locally for individual pores of a deployed FD, which leads to more accurate pressure-drop prediction. These local PM properties are assigned to the computational mesh of the FD surface in a computationally efficient manner. To demonstrate its efficacy, we model three patients' aneurysm cases treated by the Pipe Embolization Device (PED, Covidien, Irvine, CA), a widely used FD product consisting of 48 entangled wires each with a diameter of 30 $\mu m$ [2]. The FDs were virtually deployed into patient-specific aneurysms using our well-established fast virtual stenting workflow [2]. Based on the deployed FD, we generated a simplified, zero-thickness FD mesh and calculated the porous media source terms for each pore in the FD. We then applied the simplified porous media FD mesh to the computational domain of the patient-specific aneurysm and performed CFD simulations, and from the predicted flow field, calculated four hemodynamic factors: *aneurysm averaged velocity*, *inflow rate*, *shear rate*, and *the flow turnover time*. Finally, we compared the accuracy of the results and computation speed of our new PM technique with that of an explicit device simulation.

## MATERIALS AND METHODS

### *Flow Diverters as Thin-Walled and Inhomogeneous Porous Media*

The iPM approach is based on two key observations: (1) an FD is a thin layer of porous media, and (2) the pore distribution of an FD implanted in an anatomic cerebral artery is inhomogeneous.

First, as shown in Fig. 1, a flow-diverting stent is only a thin layer of metal wires, and thus the flow crossing through each pore cannot be considered a fully-developed pipe flow. For a pipe flow to become fully-developed, the pipe length must be on the order of multiples to tens of the pipe diameter, even at low Reynolds numbers [23]. However, the distance of the flow traveling through these "pipes" on the FD is less than the pore size. Due to their assumption of fully developed pipe flow, we argue that the commonly used PM methods derived from the generalized Hagen-Poiseuille equation, such as the Darcy-Forchheimer model, are unsuited for flow diverters, and will in fact overpredict the pressure drop through the device. Instead of the Darcy-Forchheimer model, our iPM approach uses the pressure-drop model across a thin wire-mesh screen based on the Idelchik equation [15] to determine the local PM properties, as detailed in the section below.

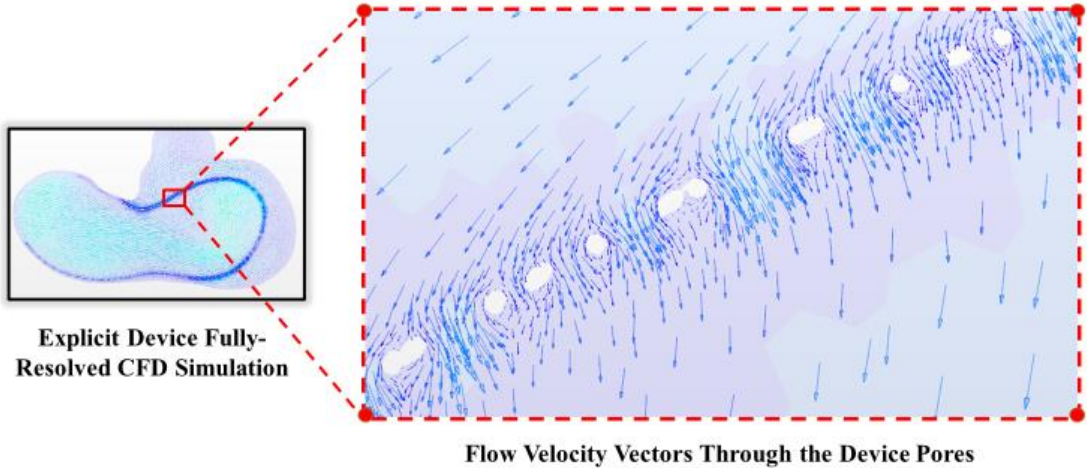

**FIGURE 1: Cross-sectional view of flow through the FD pores. Flow velocity vectors are mostly uniform in channels between the wires.**

Second, the distribution of pores in an FD that has been implanted in a patient-specific cerebral artery is inhomogeneous, with nonuniform wire spacing and angle across its surface. We aim for our iPM model to intrinsically capture this behavior. As mentioned earlier, a flow-diverting stent is woven from multiple individual wires that form a highly flexible tubular mesh. To confirm to a tortuous cerebral artery, the diamond-shaped pore cells will be squeezed or stretched to varying shapes and densities. On an inner curvature, the cells tend to be squeezed together, while on an outer curvature, cells tend to be stretched apart. Such highly non-uniform distribution of pores leads to highly non-uniform pressure drop, thus necessitating the treatment of an inhomogeneous porous media.

Without explicitly deploying the FD in the patient-specific IA, it is very difficult to predict exactly the local cell distribution. Therefore, we argue that the virtual FD deployment step should not be omitted as in the homogeneous PM approaches [8, 20]. Many fast virtual stenting methods are now available and FD deployment can be completed in seconds [2, 24-27]. Hence the FD deployment step is not the computational bottleneck. For this reason, in our iPM approach, virtual FD deployment is performed first to guide the calculation of the inhomogeneous PM parameters and enable the subsequent iPM flow simulation for patient-specific aneurysms.

### *Screen-based Local Porous Media Model*

To model the flow diverter as a region of porous media, we use the screen-based PM model by Idelchik [15]. This model relates the pressure drop ($\Delta P$) of flow passing through a *thin screen of entangled wires* to geometrical and fluidic parameters as follows:

$$\Delta P = -\left\{\frac{11\mu}{d_h}u + \frac{\rho}{2}\left[1.3(1-\beta) + \left(\frac{1}{\beta}-1\right)^2\right]u^2\right\} \quad (1)$$

where $u$ is flow velocity, $\mu$ is fluid viscosity, $\rho$ is fluid density, $\beta$ is the porosity of the medium, and $d_h$ is the hydraulic diameter [15]. In this equation, the coefficient of the first-order velocity term, $11\mu/d_h$, is known as the *viscous resistance factor*, and the coefficient of the second-order velocity term, $\rho/2[1.3(1-\beta) + (1/\beta - 1)^2]$, is known as *inertial resistance factor*.

To model the FD as a fully inhomogeneous PM, the Idelchik model coefficients (*viscous* and *inertial resistance factors*) are calculated across the entire FD device, pore by pore. In order to obtain $\beta$ and $d_h$ at each pore, we first calculated 4 geometrical parameters: the total area ($A_t$) of the device cell, which is the summation of wire area ($A_w$) and pore area ($A_p$) of the cell, and the braiding angle ($\alpha$) for each cell of deployed FD. Additionally, the FD cell centroid position ($x_c, y_c, z_c$) was extracted. This quantity is used for mapping the viscous and inertial resistance factors onto the PM surface. These parameters are shown on a sample diamond-shaped FD cell in **Error! Reference source not found.**a. In **Error! Reference source not found.**b we show the relation of the cell total area ($A_t$), wire area ($A_w$), and pore area ($A_p$). For the geometrical calculations of each cell, half the wire thickness is used ($t/2$) by following the wire centerlines (dashed lines).

By calculating these geometrical parameters, the porosity and hydraulic diameter are then determined using Eqs. (2) and (3), respectively.

$$\beta = \frac{\text{Pore area } (A_p)}{\text{Total area } (A_t)} = \frac{\text{Total area } (A_t) - \text{Wire area } (A_w)}{\text{Total area } (A_t)} \qquad (2)$$

$$d_h = A_t \sin(\alpha) \qquad (3)$$

This process is performed for every FD pore cell and is saved based on the cell centroid position.

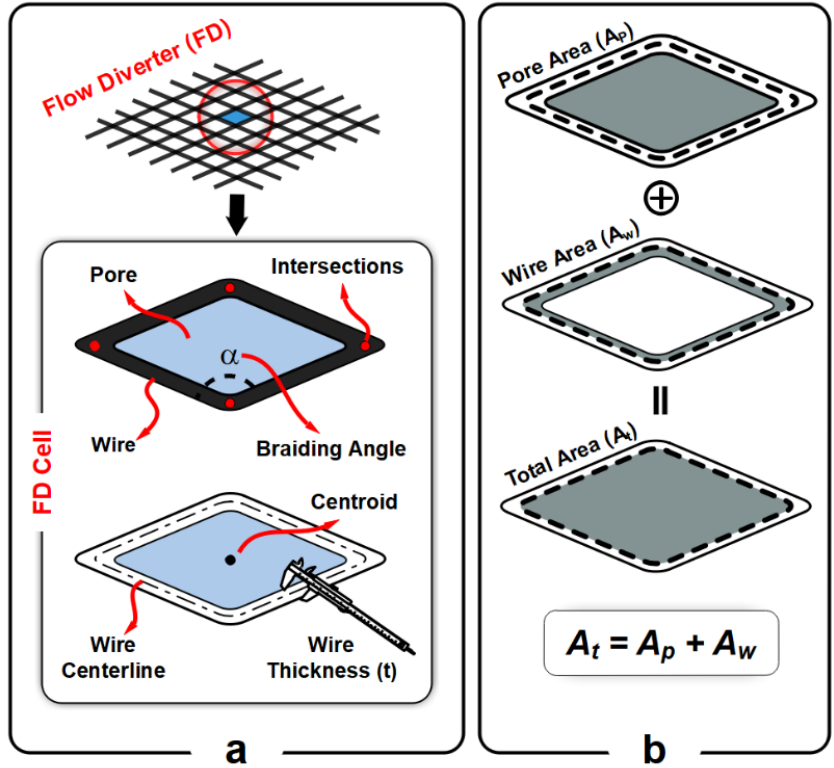

**FIGURE 2:** (a) Geometrical parameter definitions, (b) The total area ($A_t$), wire area ($A_w$), and pore area ($A_p$) relation.

### *Defining Inhomogeneous PM Properties*

In order to apply the above thin-walled local porous media model heterogeneously, we map the local PM pressure drop to the surface of a simplified zero-thickness 3D model of the implanted FD in the following four steps. We first virtually deploy the FD into the vasculature using the virtual stenting workflow of Paliwal et al. [2], acquiring an explicit 3D representation of the FD in the flow. In the second step, we extract the wire intersection points of the explicitly deployed FD and use them as vertices in a simplified, zero-thickness mesh representing the FD. Third, we use the coordinates of the wire intersections in the virtual deployment to calculate $\beta$, $d_h$, $\alpha$, and FD cell centroid. These geometrical quantities are then used to calculate the inertial and viscous resistances for each FD cell, as discussed in the above section. In the fourth step, we map the PM resistance factors for each FD cell to the simplified FD mesh by assigning them to each FD centroid using *nearest-neighbor interpolation*, as seen in FIGURE . Since the pore size is small compared to the scales of the flow, and changes between adjacent cells are small, higher order interpolation was not used. As is shown in the zoomed view, the inhomogeneous PM is represented as numerous, locally homogeneous regions with similar sizes to the FD pore size.

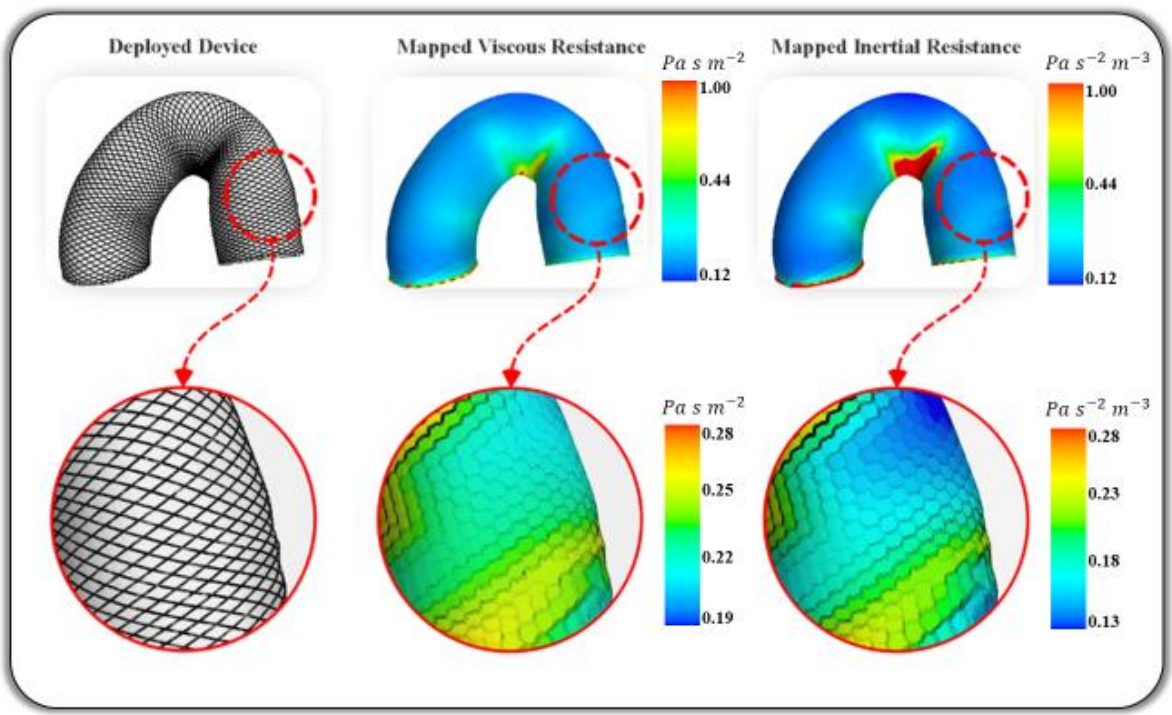

**FIGURE 3: Mapping the PM viscous and inertial resistances to the simplified flow diverter mesh using nearest-neighbor interpolation. Zoomed views (with rescaled colors) show the similarity of flow diverter pore size to the locally homogeneous PM regions.**

*Mesh Dependency*

To maximize computational speed of the simulations in this study, we performed a mesh dependency study, that is, we found the coarsest mesh size yielding outputs within 5% of simulations with an asymptotically fine mesh cell count. The mesh dependency study simulations were performed on the patient-specific aneurysm marked as Aneurysm B later in the paper using exactly the iPM workflow discussed below, with the exception of varying the mesh count and thus base mesh size. Figure 4 shows the mesh dependency of our desired quantitative hemodynamic results from iPM. Based on these results, in this study, we chose a base mesh size of 0.2mm for all simulations. Note that if only average velocity and inflow rate were desired, two post-treatment quantities found in Paliwal et al. [4] to indicate successful treatments, the cell count may be halved, thus allowing for a coarser base mesh size, boosting the computation speed.

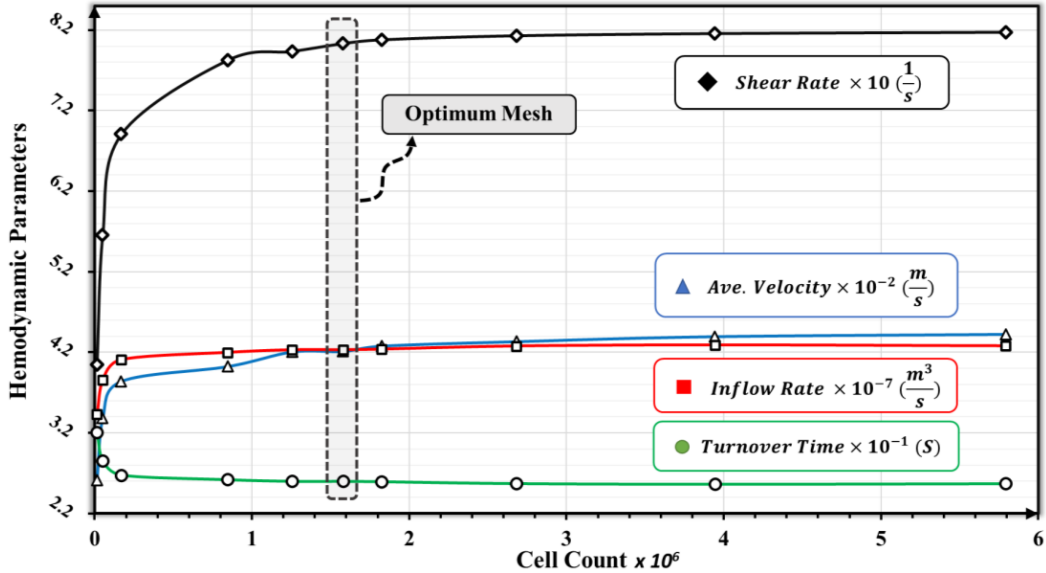

**FIGURE 4:** Mesh dependency study: Variation of aneurysm averaged velocity, shear rate, inflow rate, and turnover time as a function of cell count. The optimum base mesh size was determined to be 0.2mm.

*Testing accuracy improvement with progressively refined inhomogeneity in PM*

In order to understand the impact of adopting inhomogeneous, local PM properties in post-stenting aneurysmal flow simulation by the iPM method, we conducted a numerical experiment, *i.e.* a series of simulations for one patient-specific aneurysm treated with a FD, using progressively refined PM local properties. The entire FD region open to the aneurysm orifice is first treated as 1 homogeneous porous media, then divided into 2 equal-sized homogeneous PM regions endowed with different PM properties, then further divided it into 4 equal-sized homogeneous PM regions, and finally treated as a fully inhomogeneous PM with pore-by-pore local PM properties. Results of these four progressively refined iPM simulations were then compared against a fully-resolved simulation with the FD explicitly represented in the computational domain to evaluate their accuracy.

To do this series of numerical experiment, we first calculated the local PM parameters pore-by-pore following our new iPM strategy. To create homogeneous PM regions, we averaged the local PM properties in 1, 2, and 4 distinct regions of the aneurysm orifice, mimicking the simplifications made by heterogenization. From each simulation, we calculated the wall shear stress (WSS) at the aneurysm wall, a critical quantity for predicting aneurysm rupture propensity [1].

*iPM CFD Method Workflow*

Figure 5 shows the entire workflow that incorporates the iPM method for modeling the post-stenting hemodynamics of patient-specific IAs. The virtual device deployment step follows Paliwal et al. [2]; the iPM approach is depicted within the dashed bars, and the CFD simulation follows Paliwal et al. [4], except that the FD device is with a zero-thickness inhomogeneous PM mesh.

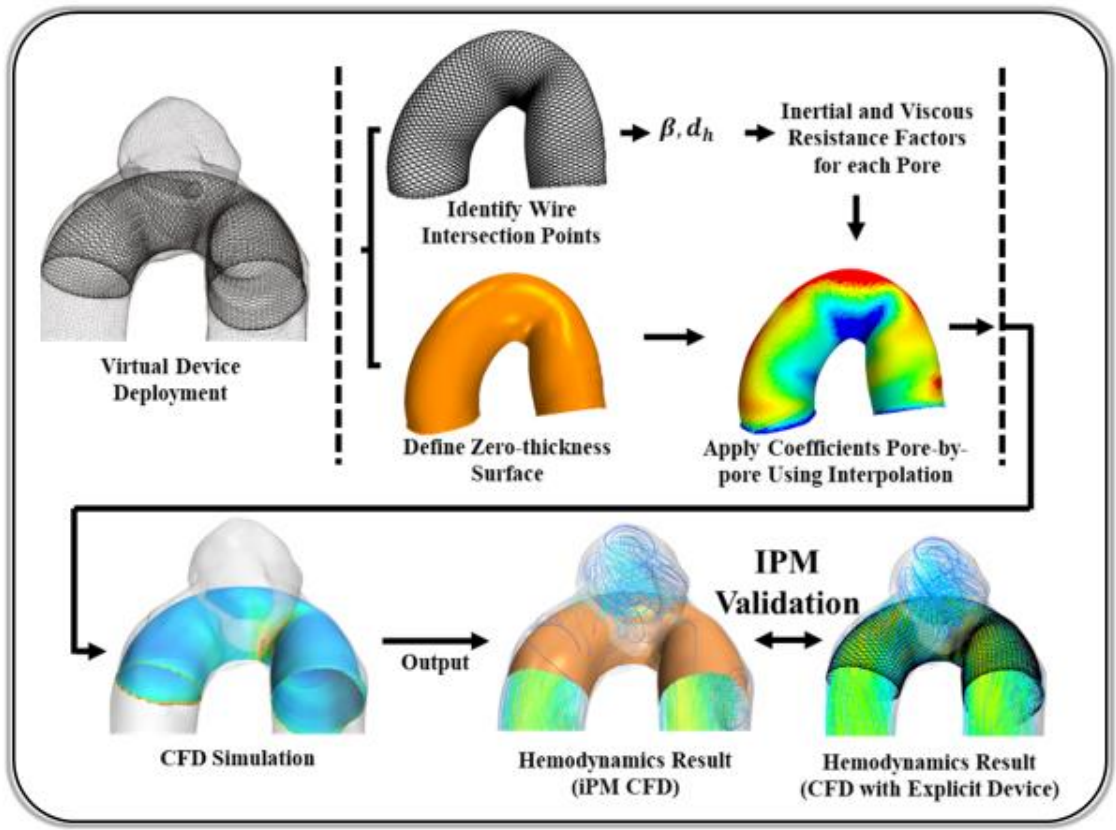

**FIGURE 5: Workflow for modeling post-treatment IA hemodynamics with a flow diverter modeled as a porous medium. The steps of the novel iPM approach are within the dashed bars.**

First, a realistic representation of a deployed FD in the vessel is generated per Paliwal et al. [2]. This fast virtual FD deployment method works by first introducing unexpanded stent along the centerline of the parent artery, then simulating the stent-surface expansion, which stops when the stent surface apposes the vessel wall. The deployed FD surface is input into the iPM step to generate a list of wire intersection points for the FD, and a zero-thickness mesh for the CFD step. First, iPM uses these wire intersections to define the PM pores and calculates the inertial and viscous resistance factors for each pore through $\beta$ and $d_h$ per Equations 2 and 3. Additionally, the virtually deployed FD is also used to generate a zero-thickness surface mesh (the orange surface in Fig. 5). This simplified zero-thickness surface mesh is imported into the simulation domain in Star-CCM+ v13.04 (CD-adapco, Melville, NY) to generate the 3D computational domain mesh using polyhedral cells with a base mesh size of 0.2mm. Nearest-neighbor interpolation is then used to apply the resistance factors for each FD pore to the zero-thickness surface mesh.

We ran the CFD simulations on the supercomputer cluster at the Center of Computational Research (CCR), University at Buffalo. All the calculations were performed on 15 nodes of 2.4 GHz Intel Xeon "Westmere" Processors Dell C6100 64-bit Linux cluster, each with 48 GB memory. Blood was assumed to be an incompressible Newtonian fluid with density of 1060 $kg/m^3$ and viscosity of 0.0035 $Pa \cdot s$. Vessel walls were considered rigid with no-slip boundary condition and the simulations were solved under steady-state conditions.

From the iPM CFD simulation results, we output hemodynamic results including flow streamlines and WSS distribution. Moreover, we calculated aneurysm average velocity and inflow rate in the aneurysm, as these quantities, which measures the post-stenting flow activities inside the aneurysm sac, may be correlated with aneurysm healing, viz. thrombotic occlusion after FD intervention. In a study of a small (N=84) cohort of patient aneurysms, these metrics have been shown to be strong predictors of failure to occlude at the 6-month follow-up [4]. In addition, we also calculated shear rate, which, at high levels, could trigger platelet activation, and average turnover time, a metric of stasis that could lead to coagulation cascade.

### *Testing the Workflow for Three Patient-specific FD-treated Aneurysms*

Due to the variations of the patient-specific arteries, the geometry of the deployed FD in individual patient-specific vasculature can vary drastically. We examined three representative internal-carotid-artery (ICA) aneurysms treated by the Pipeline Embolization Device (Medtronic), denoted as Aneurysm A, B, and C from three different patients. The vascular geometry and virtual deployment of FD are presented in Fig. 6, and aneurysm characteristics are given in Table 1. The angiographic images were retrospectively reconstructed under institutional approval. In Figure 6, the segmented 3D aneurysms with their parent arteries are shown in blue, the insets show the virtually deployed FD for each case.

To visually inspect the inhomogeneous cell distributions on the FD, we also show an *en face* view of a portion of the FD over the aneurysm neck. For the CFD simulation boundary conditions, the flow rate for each IA was scaled according to the inlet area by the factor obtained using the measured ICA flow rate in healthy subjects [28]. Based on this flow rate and inlet area, the inlet flow velocity and flow Reynolds number were calculated. The fluid boundary conditions of the CFD are listed in **Error! Reference source not found.**.

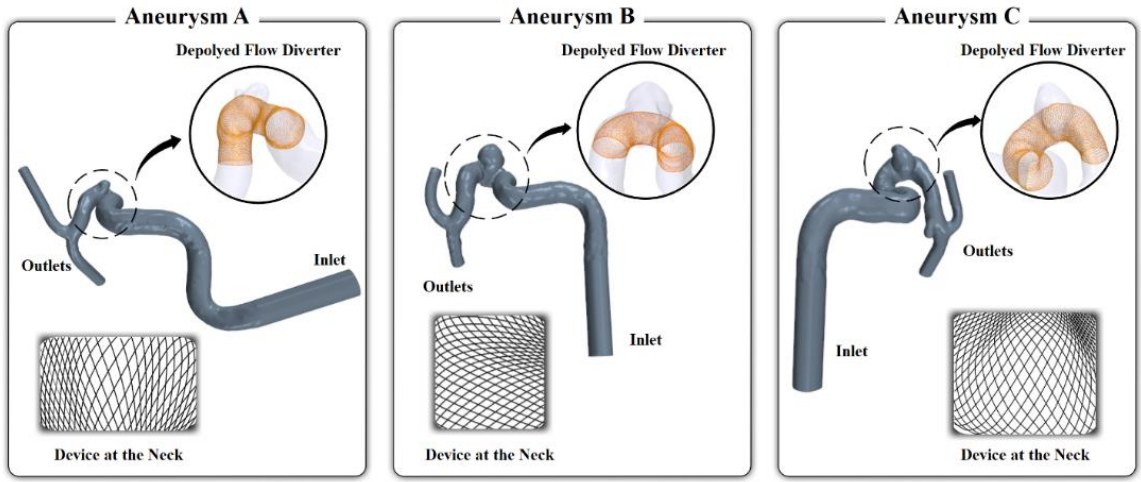

**FIGURE 6: Three representative patient-specific ICA aneurysm geometries with virtually deployed FDs.**

**TABLE 1: Patient-specific aneurysms and CFD boundary conditions**

| Cases | Aneurysm Size (mm) | Neck Diameter (mm) | Inlet Diameter (mm) | Inlet Velocity (m/s) | Inlet Reynolds Number |
|---|---|---|---|---|---|
| Aneurysm A | 2.64 | 3.46 | 5.68 | 0.152 | 260 |
| Aneurysm B | 5.51 | 5.08 | 5.22 | 0.132 | 208 |
| Aneurysm C | 3.75 | 4.32 | 5.44 | 0.141 | 231 |

*Comparison against Fully-Resolved CFD with Explicit Device*

To test and validate the new iPM porous media approach, we also ran fully resolved CFD with explicit 3D representation of the deployed FD in the three patient-specific aneurysms. Instead of turning the flow diverter into a zero-thickness mesh, the explicit 3D model of the Pipeline Embolization Device generated using the technique of Paliwal et al. [2] was imported directly in the computational domain. To resolve the flow around the stent wires, an additional mesh size constraint of 0.01mm near the FD wires was used, [29], resulting in an average of 4.5 times more mesh elements than required for the iPM approach.

# RESULTS

*Accuracy Increases with Progressively Refined Inhomogeneous PM Representation*

Figure 7 shows results of the numerical experiment on Aneurysm B with progressively refined inhomogeneous PM representations (the left 4 columns), compared against a fully resolved CFD simulation with explicit device representation as the "ground truth" (the right-most column). From left to right, the progressively refined PM simulations are ordered as follows: 1 homogeneous PM region (mimicking a homogeneous PM approach), 2 homogeneous PM regions, and 4 homogeneous PM regions, and fully heterogeneous with local pore-specific PM parameters. The top row shows the *en face* view of the virtually deployed FD in the aneurysm orifice; the bottom row shows the WSS distribution on the vessel and aneurysm wall. As the PM model was made progressively more heterogeneous, the error in the average WSS over the aneurysm sac compared to the explicit-device simulation drastically decreased from 21.6% to 3.0%. This result confirms our hypothesis, accentuating the necessity and impact of adopting local inhomogeneous PM parameters in modeling flow diverters in IAs.

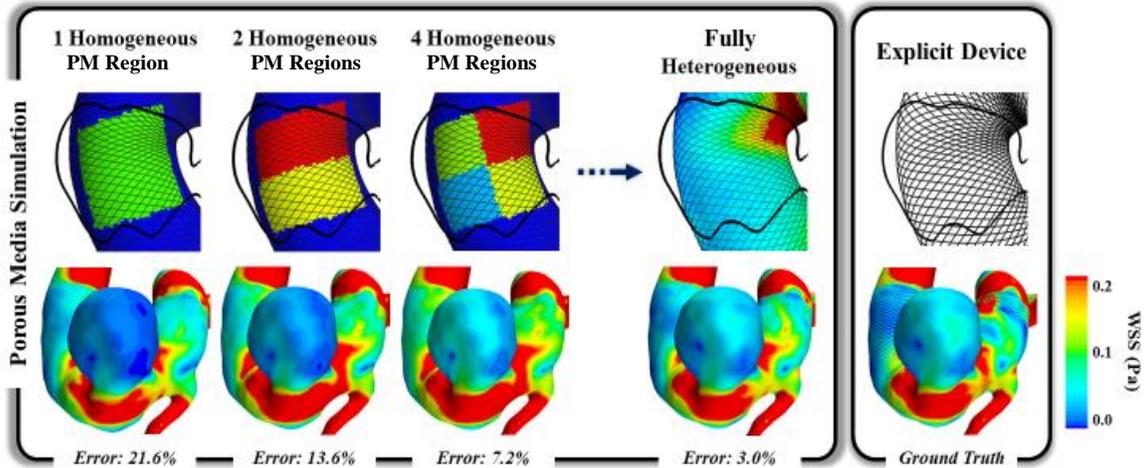

**FIGURE 7:** Illustration of the improvement in simulation predictions of hemodynamic quantities with progressively increasing inhomogeneity, compared to simulation using an explicitly represented FD. Results shown are for Aneurysm B. Top row: PM properties were averaged and applied homogeneously over 1, 2, and 4 progressively smaller homogeneous regions in the aneurysm orifice, colored red, green, blue, and cyan. The black curved outline represents the flattened projection of the aneurysm orifice. Bottom row: qualitative and quantitative comparison of WSS with progressively refined inhomogeneity.

### *iPM Captures Post-treatment Hemodynamics*

Figure 8 shows the flow streamlines and WSS results on the three patient-specific cases before treatment, iPM CFD for post-stenting, and CFD with explicit FD device for post-stenting. Visually, both the iPM and the explicit-device approaches were able to capture the flow reduction due to implanted FDs to similar degrees. In particular, the post-stenting WSS patterns are extremely similar between the iPM and explicit-device approach. We further quantified the aneurysm-averaged velocity, inflow rate, shear rate, and turnover time (Fig. 9). For all three aneurysms, both the iPM and explicit-device approaches reports similar drastic reductions of average velocity, inflow rate and shear rate, as well as similar drastic increases of fluid turnover time. The differences of iPM predictions of these quantities were 1%-6% from the explicit-device results. This is within the modeling uncertainties arising from using typical CFD assumptions for a complex-geometry intracranial aneurysm flow [29] and also within the velocity uncertainties for different, mesh-independent mesh sizes [30]. As such, we find the inhomogeneous PM approach yields reasonably accurate estimations of the hemodynamics of FD-treatment for patient-specific aneurysms.

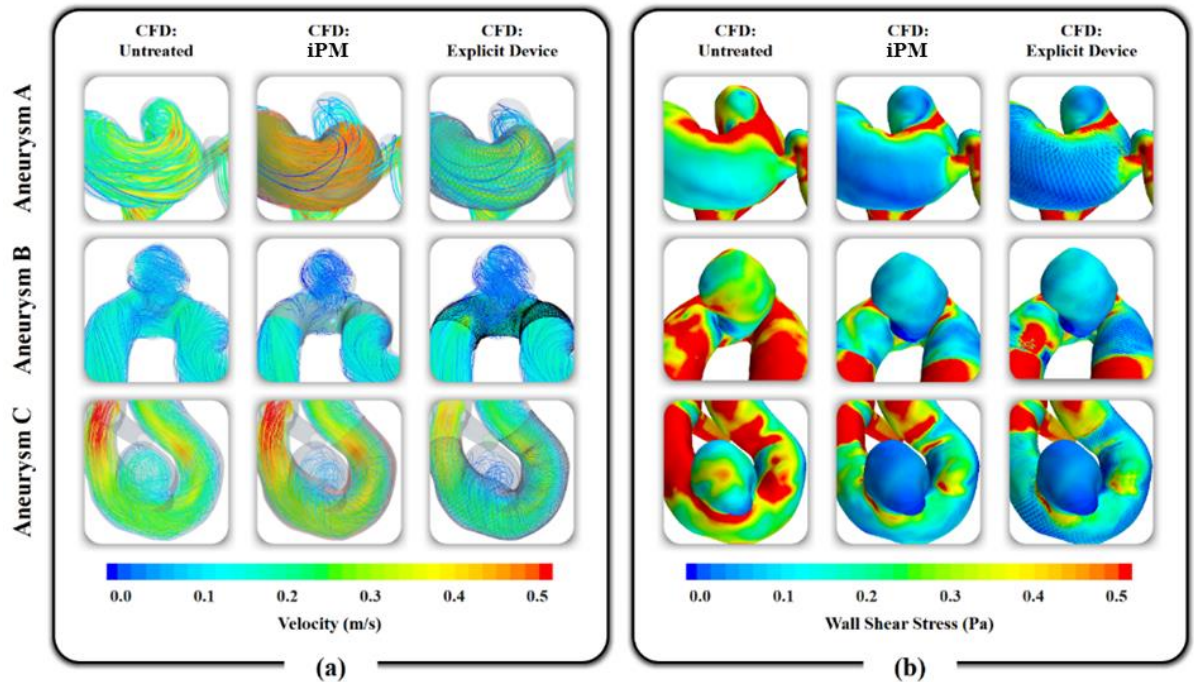

**FIGURE 8:** Qualitative comparison of (a) Flow streamlines (b) wall shear stress (WSS) for untreated (first column), treated by inhomogeneous PM (second column), and treated with explicit device (third column), in three selected cases of Aneurysms A, B, and C.

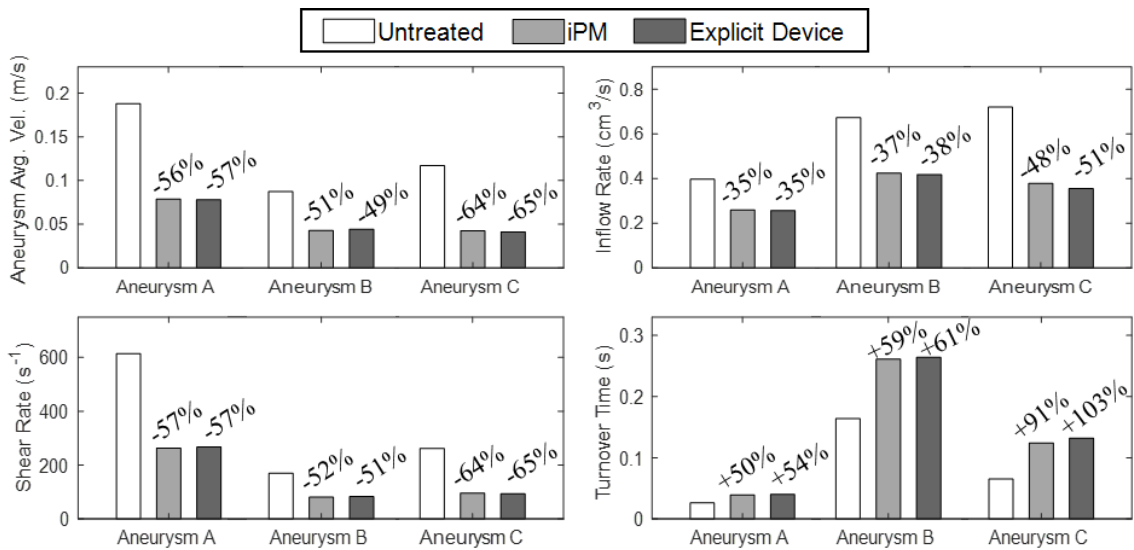

**FIGURE 9.** Quantitative comparison of aneurysm averaged velocity, inflow rate, shear rate, and turnover time in Aneurysms A, B, and C for the untreated case compared to virtual treatments using a flow diverter by iPM and an explicitly modeled FD. The percent differences of iPM and the explicit device case to the untreated case are listed above their respective bars.

## *Improvement in Computational Efficiency by Inhomogeneous PM*

The time performance and mesh sizes of the CFD simulations for the three patient-specific FD cases are listed in Table 2. We observe that across all three aneurysms, the computational time of the iPM approach was substantially shorter compared to the explicit-device CFD: the mesh counts were on average 78% smaller, and the simulation runtimes were on average 80% less. As such, the runtime of all cases decreased from hours to minutes. This means that iPM-based CFD was able to model the post-stenting hemodynamics with similar computational efficiency as the CFD in untreated aneurysms, with only approximately 5 added minutes (for adding the mesh nodes of the FD). Note that these results were acquired using steady-state flow conditions. For pulsatile flow simulations, the time savings by iPM will be even more drastic.

**TABLE 2. Untreated, Porous Media (PM), and explicit device CFD simulation comparison in terms of mesh cell count and CFD simulation runtime in Aneurysms A, B, and C. The base mesh size was 0.2mm in all cases, and the explicit device simulation had the additional constraint of 0.01mm mesh size near the FD wires.**

| Aneurysm | CFD Model | Mesh Cell Count | Simulation Runtime |
|---|---|---|---|
| A | Untreated | 1,400,203 | 14 m 03 s |
| | iPM | 1,978,101 | 21 m 51 s |
| | Explicit Device | 8,100,775 | 98 m 17 s |
| | % Improvement, Treated | 75.6% | 78.8% |
| B | Untreated | 1,277,828 | 12 m 35 s |
| | iPM | 1,579,283 | 17 m 1 s |
| | Explicit Device | 7,544,833 | 86 m 40 s |
| | % Improvement, Treated | 79.1% | 80.4% |
| C | Untreated | 1,172,290 | 9 m 42 s |
| | iPM | 1,673,854 | 12 m 41 s |
| | Explicit Device | 7,915,936 | 68 m 40 s |
| | % Improvement, Treated | 78.9% | 81.5% |

# DISCUSSION

Flow diverters have become increasingly popular for treating complex aneurysms, but the selection of device, implantation strategy and the long-term healing rate are highly patient-dependent. Persistent aneurysmal flow activities and incomplete occlusion of the aneurysm in the long run exposes the FD-treated aneurysm patient to increased risk of aneurysm rupture and leaves the patient with fewer options of re-treatment. Paliwal et al. [4] has recently shown that the 6-month patient outcome of FD treatment can be predicted based on pre- and immediately post-treatment hemodynamics (aneurysm-averaged flow velocity, inflow rate etc.), calculated by virtually implanting FDs into patient-specific aneurysm models and running CFD to obtain hemodynamics. Therefore, empowering the neurointerventionalists with computation-based, patient-specific virtual treatment and "what-if" options is vital for improving patient outcome.

While computational *accuracy* is a prerequisite for such individualized medicine, bedside tools must also be computationally *efficient* to facilitate rapid, time-sensitive clinical decision-making. In hemodynamic simulations of virtually stented IAs, the complex geometry of flow-diverting stents has been an obstacle in implementing CFD at the bedside. In this study, we have demonstrated that the computational cost of post-FD hemodynamic simulation can be drastically reduced by representing a deployed FD as a zero-thickness porous-medium tube with locally-varying PM properties. The post-stenting CFD simulation time was reduced from 1.5 hours for steady-state CFD in the presence of an explicit FD device down to 20 minutes or less by using the iPM method —similar to CFD before FD treatment, without compromising the accuracy of the hemodynamic predictions. Because the FD is not explicitly represented in the post-stenting CFD, the number of computational mesh nodes is drastically reduced. For steady-state flow simulations, the iPM results were accurate within 6% of the CFD results with explicitly-resolved devices.

Porous-media approaches have been widely used in industrial CFD, but directly importing them to represent the flow change induced by an implanted flow diverter in a patient's cerebrovascular system has two major problems: the flow diverter (a dense stent mesh) is not a thick three-dimensional porous medium, and the mesh distribution is highly inhomogeneous depending on the vessel curvature. Our novel iPM approach simultaneously incorporates two improvements over the past PM approaches attempted for representing FD, thereby more realistically modelling the FD in the flow domain. First, we model an FD as a thin wire screen described by the Idelchik model [15] instead of the classic Darcy model for 3D porous media. Second, we endow the screen with locally varying PM properties based on each individual pore in the PM. Benefiting from these two improvements, the iPM method affords a much more accurate, simplified model for CFD of the post-stenting hemodynamics compared to the prior FD approaches.

In the future, iPM computational speed can be further improved by optimizing the simulation mesh size for specific hemodynamic results. For example, if solely the aneurysm-average velocity and inflow rate are desired, the mesh may be coarsened as these quantities were mesh independent at half the mesh count (Fig. 4), boosting processing times without a loss in accuracy.

Although we have demonstrated the iPM technique for three retrospective patient cases treated by a single Pipe Embolization Device (one type of flow diverter), this computational methodology can be applied to other FD products, stents that are not FDs, and for multiple layers of FDs or stents, allowing clinicians more choices in simulated treatment options. This is possible because regardless of the type or number of FDs, the PM pore size is much larger than the device wire thickness, and as such, the wire intersections of even different FDs layered on top of one another may be approximated to be on a single zero-thickness plane. In this case, the definitions of Fig. 2 hold, but the pores may no longer be diamond-shaped. For example, one type of FD called the Flow Re-direction Endoluminal Device (FRED) have additional wires that produce more complex pore shapes than simple diamonds. For each pore regardless of shape, $d_h$ and $\beta$ in the Idelchik model may still be calculated and implemented; however the forms may be more complex than that of Eqs. (2) and (3), depending on the pore shape.

In conclusion, iPM is a fast and relatively accurate computational tool for modeling the aneurysmal flow induced by endovascular FDs. It may be a potential candidate for implementation in the clinical setting.

## ACKNOWLEDGEMENTS

This work was supported by the National Institutes of Health grant R01NS091075 (HM). We thank valuable technical input by Nikhil Paliwal and acknowledge the Center for Computational Research at the University at Buffalo for providing the computational support for CFD simulations.

# REFERENCES

1. Meng, H., et al., *High WSS or low WSS? Complex interactions of hemodynamics with intracranial aneurysm initiation, growth, and rupture: toward a unifying hypothesis.* American Journal of Neuroradiology, 2014. **35**(7): p. 1254-1262.
2. Paliwal, N., et al., *Virtual stenting workflow with vessel-specific initialization and adaptive expansion for neurovascular stents and flow diverters.* Computer methods in biomechanics and biomedical engineering, 2016. **19**(13): p. 1423-1431.
3. Biondi, A., et al., *Neuroform stent-assisted coil embolization of wide-neck intracranial aneurysms: strategies in stent deployment and midterm follow-up.* Neurosurgery, 2007. **61**(3): p. 460-469.
4. Paliwal, N., et al., *Outcome prediction of intracranial aneurysm treatment by flow diverters using machine learning.* Neurosurgical Focus, 2018. **45**(5): p. E7.
5. Damiano, R.J., et al., *Improving accuracy for finite element modeling of endovascular coiling of intracranial aneurysm.* 2019. **14**(12): p. e0226421.
6. Yadollahi-Farsani, H., et al., *Numerical study of hemodynamics in brain aneurysms treated with flow diverter stents using porous medium theory.* Comput Methods Biomech Biomed Engin, 2019: p. 1-11.
7. Li, Y., et al. *Modelling flow-diverting stent as porous medium with different permeabilities in the treatment of intracranial aneurysms: a comparison of a successfully treated case and an unsuccessful one.* in *The 8th International Conference on Computational Methods (ICCM2017).* 2017.
8. Augsburger, L., et al., *Intracranial stents being modeled as a porous medium: flow simulation in stented cerebral aneurysms.* Ann Biomed Eng, 2011. **39**(2): p. 850-63.
9. Karmonik, C., et al., *Quantitative comparison of hemodynamic parameters from steady and transient CFD simulations in cerebral aneurysms with focus on the aneurysm ostium.* 2015. **7**(5): p. 367-372.
10. Löhner, R., S. Appanaboyina, and J.R. Cebral, *Comparison of body-fitted, embedded and immersed solutions of low Reynolds-number 3-D incompressible flows.* International Journal for Numerical Methods in Fluids, 2008. **57**(1): p. 13-30.
11. Whitaker, S., *The method of volume averaging.* Vol. 13. 2013: Springer Science & Business Media.
12. Huang, H. and J.A. Ayoub. *Applicability of the Forchheimer equation for non-Darcy flow in porous media.* in *SPE Annual Technical Conference and Exhibition.* 2006. Society of Petroleum Engineers.
13. Valdes-Parada, F.J., et al., *Validity of the permeability Carman–Kozeny equation: A volume averaging approach.* 2009. **388**(6): p. 789-798.
14. Raschi, M., et al., *Strategy for modeling flow diverters in cerebral aneurysms as a porous medium.* Int J Numer Method Biomed Eng, 2014. **30**(9): p. 909-25.
15. Idelchik, I.E., *Handbook of hydraulic resistance.* Washington, DC, Hemisphere Publishing Corp., 1986, 662 p. Translation., 1986.
16. Kim, M., et al., *Comparison of two stents in modifying cerebral aneurysm hemodynamics.* 2008. **36**(5): p. 726-741.
17. Dazeo, N., et al., *A comparative study of porous medium CFD models for flow diverter stents: Advantages and shortcomings.* Int J Numer Method Biomed Eng, 2018. **34**(12): p. e3145.
18. Romero, E., et al., *Flow diverter stents simulation with CFD: porous media modelling.* 2017. **10160**: p. 101601F.
19. Ohta, M., et al., *Parametric study of porous media as substitutes for flow-diverter stent.* Biomaterials and Biomedical Engineering, 2015. **2**(2): p. 111-125.
20. Li, Y., et al., *Numerical simulation of aneurysmal haemodynamics with calibrated porous-medium models of flow-diverting stents.* J Biomech, 2018. **80**: p. 88-94.
21. Carman, P.C., *Flow of gases through porous media.* 1956.
22. Peric, M. and S.J.D. Ferguson, *The advantage of polyhedral meshes.* 2005. **24**: p. 45.

23. Munson, B.R., et al., *Fluid mechanics*. 2013: Wiley Singapore.
24. Zhang, Q., et al., *Phantom-based experimental validation of fast virtual deployment of self-expandable stents for cerebral aneurysms.* BioMedical Engineering OnLine, 2016. **15**(2): p. 431-437.
25. Larrabide, I., et al., *Fast virtual deployment of self-expandable stents: method and in vitro evaluation for intracranial aneurysmal stenting.* Medical image analysis, 2012. **16**(3): p. 721-730.
26. Zhang, Q., et al., *Efficient simulation of a low-profile visualized intraluminal support device: a novel fast virtual stenting technique.* Chinese Neurosurgical Journal, 2018. **4**(1): p. 1-7.
27. Zhong, J., et al., *Fast virtual stenting with active contour models in intracranical aneurysm.* Scientific reports, 2016. **6**(1): p. 1-9.
28. Paliwal, N., et al. *Fast virtual stenting with vessel-specific initialization and collision detection*. in *ASME 2014 International design engineering technical conferences and computers and information in engineering conference*. 2014. American Society of Mechanical Engineers Digital Collection.
29. Paliwal, N., et al., *Methodology for computational fluid dynamic validation for medical use: application to intracranial aneurysm.* Journal of biomechanical engineering, 2017. **139**(12).
30. Spiegel, M., et al., *Tetrahedral vs. polyhedral mesh size evaluation on flow velocity and wall shear stress for cerebral hemodynamic simulation.* Computer methods in biomechanics and biomedical engineering, 2011. **14**(01): p. 9-22.